\documentclass[onecolum,secnumarabic,amssymb, nobibnotes,longbibliography,preprint]{revtex4-1}

\usepackage{graphicx,color,amsmath}
\usepackage[normalem]{ulem}

\begin{document}
	%\linenumbers
	
	\title{Diffusion in Phase Space as a Tool to Assess Variability of Vertical Centre-of-Mass Motion During Long-Range Walking }
	
	\author{N. Boulanger$^1$}
	\email[e-mail:]{nicolas.boulanger@umons.ac.be}
	\author{F. Buisseret$^{2,3}$}
	\email[e-mail:]{buisseretf@helha.be}
	\author{V. Dehouck$^{1,4}$}
	\email[e-mail:]{victor.dehouck@alumni.umons.ac.be}
	\author{F. Dierick$^{2,5,6}$}
	\email[e-mail:]{frederic.dierick@gmail.com}
	\author{O. White$^4$}
	\email[e-mail:]{olivier.white@u-bourgogne.fr}
	
	\affiliation{$^1$ Service de Physique de l'Univers, Champs et Gravitation, Universit\'{e} de Mons, UMONS  Research Institute for Complex Systems, Place du Parc 20, 7000 Mons, Belgium }
	\affiliation{$^2$ CeREF, Chaussée de Binche 159, 7000 Mons, Belgium}
	\affiliation{$^3$ Service de Physique Nucl\'{e}aire et Subnucl\'{e}aire, Universit\'{e} de Mons, UMONS Research Institute for Complex Systems, 20 Place du Parc, 7000 Mons, Belgium}
	\affiliation{$^4$ Universit\'{e} de Bourgogne INSERM-U1093 Cognition, Action, and Sensorimotor Plasticity, Campus Universitaire, BP 27877, 21078 Dijon, France}
	\affiliation{$^5$ Facult\'{e} des Sciences de la Motricit\'e, Universit\'e catholique de Louvain, 1 Place Pierre de Coubertin, 1348 Louvain-la-Neuve, Belgium}
	\affiliation{$^6$ Centre National de R\'{e}\'{e}ducation Fonctionnelle et de R\'{e}adaptation -- Rehazenter, Laboratoire d'Analyse du Mouvement et de la Posture (LAMP), Luxembourg, Grand-Duch\'{e} de Luxembourg}
	
	\date{\today}%

\begin{abstract}
	When a Hamiltonian system undergoes a stochastic, time-dependent anharmonic perturbation, the values of its adiabatic invariants as a function of time follow a distribution whose shape obeys a Fokker-Planck equation. 
The effective dynamics of the body's centre-of-mass during human walking is 
expected to represent such a stochastically perturbed dynamical system. 
By studying, in phase space, the vertical motion of the body's centre-of-mass of 25 
healthy participants walking for 10-minutes at spontaneous speed, 
we show that the distribution of the adiabatic invariant is compatible with the solution of 
a Fokker-Planck equation with constant diffusion coefficient. The latter 
distribution appears to be a promising new tool for studying the long-range kinematic variability of walking.
\end{abstract}

\maketitle

\section{Introduction}

Action-angle coordinates $(I_\alpha,\theta^\alpha)$, with $\alpha=1,\dots,n\,$, 
are of central importance in the study of deterministic classical systems with finitely many degrees of 
freedom. 
A time-independent integrable Hamiltonian may indeed be formulated as a separable function 
of the action variables only: $H= \sum^n_{i=\alpha}H_{0\alpha}(I_\alpha)\,$. 
The equations of motion for such a system read \cite[Ch. 45]{L&L}: 
\begin{equation}
	\dot I_\alpha=-\frac{\partial H}{\partial \theta^\alpha}=0,\quad \dot \theta^\alpha=\frac{\partial H}{\partial I_\alpha}=:\omega_\alpha\; .
\end{equation} 
The action variables are constants of the motion and therefore $\omega_\alpha$ is also constant, 
implying that the angle coordinates read $\theta^\alpha=\omega_\alpha\, t+\theta^\alpha_0\,$ 
where  $\omega_\alpha=\frac{2\pi}{T_\alpha}\,$ and $T_\alpha$ is the period of the motion in 
the plane $(I_\alpha,\theta^\alpha)$.  Since the Kolmogorov-Arnold-Moser theorem (see 
\cite{Kolmogorov:430016,Arnold1963PROOFOA,Moser:430015} and \cite{Dumas} 
for a historical overview) and the work of Nekhoroshev 
\cite{nekhoroshev1971behavior,nekhoroshev1977exponential}, 
the action-angle variables have proven to be the most useful for the study of 
stability of dynamical systems, including chaotic systems. 
We now restrict our formalism to systems with $n=1$, whose sole degrees of freedom 
consist in the pair $(I,\theta)\,$.

Suppose that $H$ depends on a function $\lambda(t)\,$. 
The action variable $I$ then becomes time-dependent and is called an adiabatic invariant. 
On the one hand, if $\lambda$ changes slowly during the typical period of a cycle, 
then the adiabatic invariant also changes slowly: $\dot I\sim \dot \lambda$ \cite{L&L,henrard,jose}. 
On the other hand, if $\lambda$ is a perturbative stochastic noise, the adiabatic invariant 
also becomes randomly time-dependent and the deviation from its average value remains 
perturbative. 
Detailed demonstrations and bounds for the deviation may be found in 
\cite{Hasminski1966OnSP,Cogburn}. Moreover, for a Hamiltonian $H(I,\lambda(t))$ with perturbative 
stochastic noise $\lambda(t)$, it has been shown that the density $\rho(I,t)$ of the values of the 
adiabatic invariant as a function of time obeys a Fokker-Planck equation 
\cite{Bazzani94,Bazzani95,Bazzani_1998}. The latter phenomenon is a diffusion process in 
phase space. Besides its intrinsic interest, such a formalism has already found an important 
application in plasma physics, where it allows to relax standard simplifying assumptions and 
describe the problem in a less model-dependent way \cite{PhysRevLett.104.235001}. 
The Fokker-Planck equation has also been recently applied to the study of robustness in 
gene expression \cite{fkbiol}. 

Biomechanical models of voluntary rhythmic movements in humans (of which walking has been 
studied most extensively) may also benefit from the above results. Such movements are 
quasi-periodic because of physiological noise, which prevents an individual from being in the 
same invariant state during repeated movements. The resulting variability has motivated many studies of 
human gait, most of which rely on the computation of nonlinear indices to assess variability 
(Hurst exponent, fractal dimension, etc.). We refer the interested reader to the pioneering works 
\cite{Hausdorff:1995,Hausdorff:1997} and to \cite{Stergiou,Ravi} for recent reviews. 
To our knowledge, the variability of gait has never been studied by assessing the shape and time 
evolution of the distribution $\rho(I,t)$. In the present work, we will show that that the distribution 
$\rho(I,t)$ in human walking indeed obeys a Fokker-Planck equation,  
i.e. that diffusion in phase space is experimentally observable in walking. 
Biomechanical models can then inherit the advantages of this formalism.

Our paper is structured as follows. In section \ref{diffusion}, diffusion in phase space 
and its use in modelling human walking is proposed. Then, in section \ref{expe}, 
the experimental setup is presented and numerical results are given in section \ref{sec:res}. 
Finally, in section \ref{conclu}, the results are discussed and concluding remarks are given.

\section{Diffusion in phase space}\label{diffusion}
\subsection{Generalities}

Let us consider a one-dimensional Hamiltonian $H_0(I)$, where $I$ and $\theta$ are the action and 
angle coordinates, respectively.
Suppose that a time-dependent stochastic perturbation is added to $H_0$ and that the 
latter Hamiltonian satisfies the stability assumptions underlying the Nekhoroshev theorem 
\cite{nekhoroshev1971behavior,nekhoroshev1977exponential}. 
The total Hamiltonian $H$ may be written as follows: 
\begin{equation}\label{Hdef}
H= H_0(I) + \epsilon\ \xi(t)\ {\cal V}(I,\theta)\;,
\end{equation} 
 where $0<\epsilon\ll 1$, and where $\xi(t)$ is a stochastic noise with vanishing mean value. 
 Under the dynamics controlled by $H$, the action variable becomes time-dependent and the deviation 
 from the initial value $I_0$ is of order $\sqrt{\epsilon}\,$ up to a time of order $1/\epsilon\,$ 
 or even better \cite{Hasminski1966OnSP,Cogburn}. 
 More precisely, $\vert I(t)-I_0\vert=O(\sqrt \epsilon)$ and a time-dependent 
 density distribution $\rho(I,t)$ of the values of the adiabatic invariant
can be associated with its time evolution $I(t)\,$.
As shown and illustrated in \cite{Bazzani94,Bazzani95,Bazzani_1998}, 
the density distribution $\rho(I,t)$ obeys a particular  Fokker-Planck equation given by
 \begin{equation}\label{FP0}
 	\partial_t\rho=\partial_ I ( D(I)\partial_I\rho )\;,
 \end{equation}
where the function $D(I)\,$ is called the diffusion function.
Considering the Fourier decomposition ${\cal V}(I,\theta)=\sum_k{\cal V}_k(I)\, {\rm e}^{i k \theta}$
of the perturbation function that appears in the Hamiltonian, the following expression 
is obtained \cite{Bazzani_1998} for the diffusion function:
 \begin{equation}\label{FP1}
 D(I)=\frac{\epsilon^2}{2}\sum_k k^2 \vert {\cal V}_k(I)\vert^2 \tilde\phi(k\omega)\;,
\end{equation}
where $\tilde\phi(\nu)$ is the noise spectral density, 
i.e. $\tilde\phi(\nu)=\int^{+\infty}_{-\infty}\phi(u)\cos(\nu u)\, du$ 
with the autocorrelation function
\begin{equation}
	\phi(u)=\lim_{T\to+\infty}\frac{1}{T}\int^{T}_0\xi(t)\xi(t+u)\, du \;.
\end{equation}
Two particular cases can be highlighted. 
First, when $H=(\omega +\epsilon\, \xi(t) ) I\,$, 
only the $k=0$ mode ${\cal V}_0$ is nonzero and $D=0$. There is no diffusion in a pure harmonic 
oscillator with randomly perturbed frequency \cite{Bazzani94}. Second, in the case of constant 
diffusion coefficient, the normalised solution of (\ref{FP0}) on the interval $I\in[0,+\infty[$ with boundary 
conditions $\rho(I,0)=\delta(I-I_0)\, \Theta(t)\,$, $\Theta$ being the Heaviside step function, 
and  $\rho(0,t)=0=\lim_{I\to+\infty}\rho(I,t)$, may be obtained:   
\begin{equation}\label{solu}
	\rho(I,t)=\Theta(t)\ \frac{{\rm e}^{-\frac{(I-I_0)^2}{4Dt}}
	-{\rm e}^{-\frac{(I+I_0)^2}{4Dt}}}{\sqrt{4\pi D t}\ {\rm erf}\left(\frac{I_0}{\sqrt{4Dt}} \right) }
	=\Theta(t)\ \frac{{\rm e}^{-\frac{(I-I_0)^2}{4Dt}}}{\sqrt{4\pi D t}} \ 
	\frac{1-{\rm e}^{-\frac{I\, I_0}{Dt}}}{{\rm erf}\left(\frac{I_0}{\sqrt{4Dt}} \right)}\;.
\end{equation}
The normalisation is such that $\int^{+\infty}_0\rho(I,t)\, dI=\Theta(t)$.
From now on, we will be interested in the second case of a constant but non-vanishing 
diffusion fonction.

We refer the interested reader to \cite{risken1989fpe} for a general discussion of the construction of 
solutions of the Fokker-Planck equation, and to \cite{PhysRevE.72.020101,LIN2012386} for explicit 
solutions with non-constant and nonzero diffusion and drift coefficients.

\subsection{Application to human walking}

It is known that the vertical displacement of the body's centre-of-mass (COM) during human 
bipedal walking at spontaneous speed is compatible with a simple, spring-mass-like, model, 
see for example the famous work \cite{cavagna}. 
It is therefore tempting to model the vertical motion of the COM by the 
harmonic oscillator Hamiltonian 
$H_0=\frac{1}{2}(P^2+\omega^2Q^2)=\omega\, I\,$, where $P$ and $Q$ are the 
vertical momentum and position of the COM, respectively. 
By definition, and assuming the standard relation $P\propto \dot Q\,$, one has
\begin{equation}
	I=\frac{1}{2\pi}\oint_{\Gamma} P\, dQ =\frac{T\overline{E_c}}{\pi}\;,
\end{equation} 
with $\Gamma$ a cycle in phase space, $T$ the duration of the cycle
and $\overline{E_c}$ the averaged kinetic energy over~$\Gamma\,$.

Some phenomena suggest that the inclusion of other terms, at least in the perturbation, 
is necessary to obtain a more realistic model. First, the minimum (maximum) of the potential energy 
and the maximum (minimum) of kinetic energy are not reached at exactly the same time: 
a time shift of about 3 \% of the gait cycle duration is observed \cite{Cavagna2020}. 
Such a feature requires a time-dependent correction to be added. Second, the Hamiltonian $H_0$ 
corresponds to a linearised pendulum only in the limiting case of small amplitudes.
Anharmonic corrections should be added. The interested reader will find in \cite{Whittington} 
a more explicit model of the pendulum in which the potential term is nonlinear, 
and in \cite{BRIZARD2013511} a computation of action-angle variables for the fully 
non-linear pendulum with Hamiltonian $H_0=\frac{P^2}{2}+1-\cos Q$. 
Third, the parameters of the model ($\omega$ in our case) must have some 
time-dependent variability due to physiological noise; 
the state of a complex system like the human body is not identical from one gait cycle to another. 

In view of the above discussion, a Hamiltonian of the form (\ref{Hdef}) in which $H_0$ is not 
a pure harmonic oscillator seems to be a relevant model of the vertical COM dynamics in 
action-angle formalism. As far as the perturbation $H_1(I,\theta)=\epsilon\ \xi(t)\ {\cal V}(I,\theta)$ 
is concerned, 
we will consider the simplest nontrivial ansatz with a  
constant but non-vanishing diffusion coefficient $D\,$. Referring to (\ref{FP1}), this 
implies that all the functions ${\cal V}_k(I)$ are constant so that $H_1(I,\theta)$ 
only depends on the angle variable $\theta\,$.
It does not depend on the total amount of action or energy in the system 
but only on time through the stochastic noise $\xi(t)$ and on the position in the 
cycle through ${\cal V}(\theta)\,$. Therefore, we assume that the influence of 
physiological noise on walking is related to the position in the gait cycle and not 
to the total action or the averaged kinetic energy of the walker -- recall that $I\sim \overline{E_c}$.
Consequently, Eq. (\ref{FP0})  with a nonzero diffusion coefficient yields the heat
equation 
 \begin{equation}\label{FP0Bis}
 	\partial_t\rho = D\,\partial^2_I \rho \;,
 \end{equation}
and 
the diffusion of the adiabatic invariant should be observable experimentally.

\section{Experimental setup}\label{expe}
 \subsection{Protocol}
The protocol was validated by the Academic Ethical Committee Brussels Alliance for Research 
and Higher Education (B200-2021-123). Participants were healthy students recruited in the 
physiotherapy department of the Haute-Ecole Louvain en Hainaut (Montignies-sur-Sambre, Belgium). 
After being informed about the study, each participant signed an informed consent form. 

Biometric data were first collected (age, weight, height), as well as information on the wearing of 
orthopaedic insoles and the participant's medical and trauma history. The participant is then asked to 
put on a tight-fitting garment. In order for his or her movements to be recorded by a Vicon 
optoelectronic system (Vicon Motion Systems Ltd, Oxford Metrics, Oxford, UK) consisting of 8 cameras 
(Vero v.2 .2) with a recording frequency of 120 Hz, 4 reflective markers with a diameter of 14 mm were 
placed on the participant according the Plug-In-Gait model (Oxford Metrics, Oxford, United Kingdom): 
Left Anterior Superior Iliac Spine [LASI], Right Anterior Superior Iliac Spine [RASI], Left Posterior 
Superior Iliac Spine [LPSI], and Left Posterior Superior Iliac Spine [RPSI].

After this preparatory phase, the participant walked for 3 minutes on an N-Mill instrumented treadmill 
(Motekforce Link, The Netherlands). The purpose of this familiarisation phase is to determine the 
participant's spontaneous walking speed. No other data were recorded during this period. After the 
walking speed was recorded, the participant walked on the treadmill for 10 minutes at the previously 
determined spontaneous speed. During these 10 minutes, the average number of steps per minute was 
measured by the treadmill and the three-dimensional trajectory of the 4 markers, $\vec x_a(t)$, was 
recorded by the Vicon system using the Vicon Nexus software (v.2.7.1, Oxford Metrics, Oxford, UK).

The general characteristics of our participants are listed in Table \ref{tab1}. We note that an initial 
analysis of these data was presented in a recent work \cite{AdiabaticGait}, in which we showed that an 
adiabatic invariant exists in the vertical motion of the COM. Here we go further in the analysis to assess 
whether or not the variability of the latter adiabatic invariant is modelled by Eq. (\ref{FP0Bis}).

\begin{table}[h]
	\caption{Features of our population. Results are written under the form mean$\pm$ SD. The number of gait cycles performed in 10 minutes by the participants is given under the form median [Q1-Q3].}
	\centering
	%% \tablesize{} %% You can specify the fontsize here, e.g., \tablesize{\footnotesize}. If commented out \small will be used.
	\begin{tabular}{cc}
\hline
		Participants ($n$) & 	25 \\
		Age (years) &	23 [20$-$23] \\
		Mass (kg)&	65.0 [58.8$-$73.4] \\
		Height (cm)&	169  [164$-$176] \\
		Walking speed (km/h) & 3.9 [3.5$-$4.2]\\
		Sex (men/women) &	9/16 \\
		Gait cycles  &	532 [513-552] \\
\hline
	\end{tabular}\label{tab1}
\end{table}

 \subsection{Data processing}
 
For a given participant, the position of the centre-of-mass is defined as the average position of the four 
markers: $\vec x_{COM}=\sum_a\vec x_a/4$. We focus here on the vertical component of the COM 
motion, $Q(t)$. To reduce measurement artefacts, $Q(t)$ was filtered with a fourth-order Butterworth 
low-pass filter, preserving 99.99\% of the signal power. Cubic spline interpolation of the data was also 
performed, multiplying the frequency by 10 to 1200 Hz. The speed $\dot Q$ is computed from the time 
series $Q$ by finite differentiation. 

An identification $P=\dot Q$ is performed, i.e., we assume standard Hamiltonian dynamics and set 
the mass scale equal to 1 (this normalisation removes the variability induced by participants' masses). 
We then identify gait cycles by analysing the peaks in $Q(t)$: The duration of gait cycles $i$, 
$T_i=t_{i+2}-t_i$, were computed from the times $t_i$ at which the peaks occur. The times $t_i$ may 
be defined as the times at which a new step begin, a gait cycle consisting in two steps (left and right). 
Then the average kinetic energies, $\overline{E_c}_i\,$, were computed as the mean values of $\dot 
Q^2/2$ on the successive cycles, and the adiabatic invariants
\begin{equation}
	I_i=\frac{T_i\overline{E_c}_i}{\pi}
\end{equation} 
were also computed.
 
The values collected in the sets $A_i=\{I_{j\leq i}\}$ are then binned according to Sturges rule 
\cite{Sturges}, leading to $n$ bins. The centres $I^{(i)}_a$ and frequencies $\varphi^{(i)}_a$, i.e. the number of items in bin $a$ divided by total number of items, are computed, with $a=1,\dots,n$. The experimentally computed distribution $\rho^{{\rm exp}}(t_i,I)$ of the 
adiabatic invariant after a walking duration $t_i$ is defined via 
$\rho^{{\rm exp}}(t_i,I)=\left(I^{(i)}_a,\varphi^{(i)}_a\right)$.

A fit of the form (\ref{solu}) 
is then performed on the sets $\rho^{{\rm exp}}(t_{i\geq 100},I)$ using the 
least-squares method and the parameters $I_{0i}$ and $D_i$ are recorded. The latter parameters are the fitted values of $I_0$ and $D$ at time $t_i$. No fit was made for the first 100 points. This threshold is arbitrary but avoids situations where the distribution has too little structure for the adjustment to be relevant. Finally, we compute the average values $I_0={\bm E}(I_{0i})$ and $D={\bm E}(D_i)$, resulting in a distribution  (\ref{solu}) called the model, $\rho^{{\rm th}}(t,I)$. 
%The drift of the distribution is estimated as follows. The slopes $\delta_i$ of $I_{0i}$ versus $t$ are computed by linear regression. Then we define the drift as $\delta=100 E(\delta_i T_i)/I_0$, where $T_i$ are the time periods for which the regressions are performed. 

The compatibility of the experimental distributions $\rho^{{\rm exp}}(t_{i\geq 100},I)$ and the model predictions 
 $\rho^{{\rm th}}(t_i,I)$ is assessed by a Kolmogorov-Smirnov test with a significance level 0.05. We note $\Pi$, the percentage of tests with $p>0.05$, i.e., the percentage of cases in which the model is incompatible with the experimental data. One-sample t-tests were performed with null hypothesis of zero mean for $I_0$ and $D$.

All the above computations were performed using the free software R (v. 4.1.0, https://www.r-project.org).

%%%%%%%%%%%%%%%%%%%%%%%%%%%%%%%%%%%%%%%%%%
\section{Results}\label{sec:res}

\begin{figure}[h]
	\centering
	\includegraphics[width=14cm]{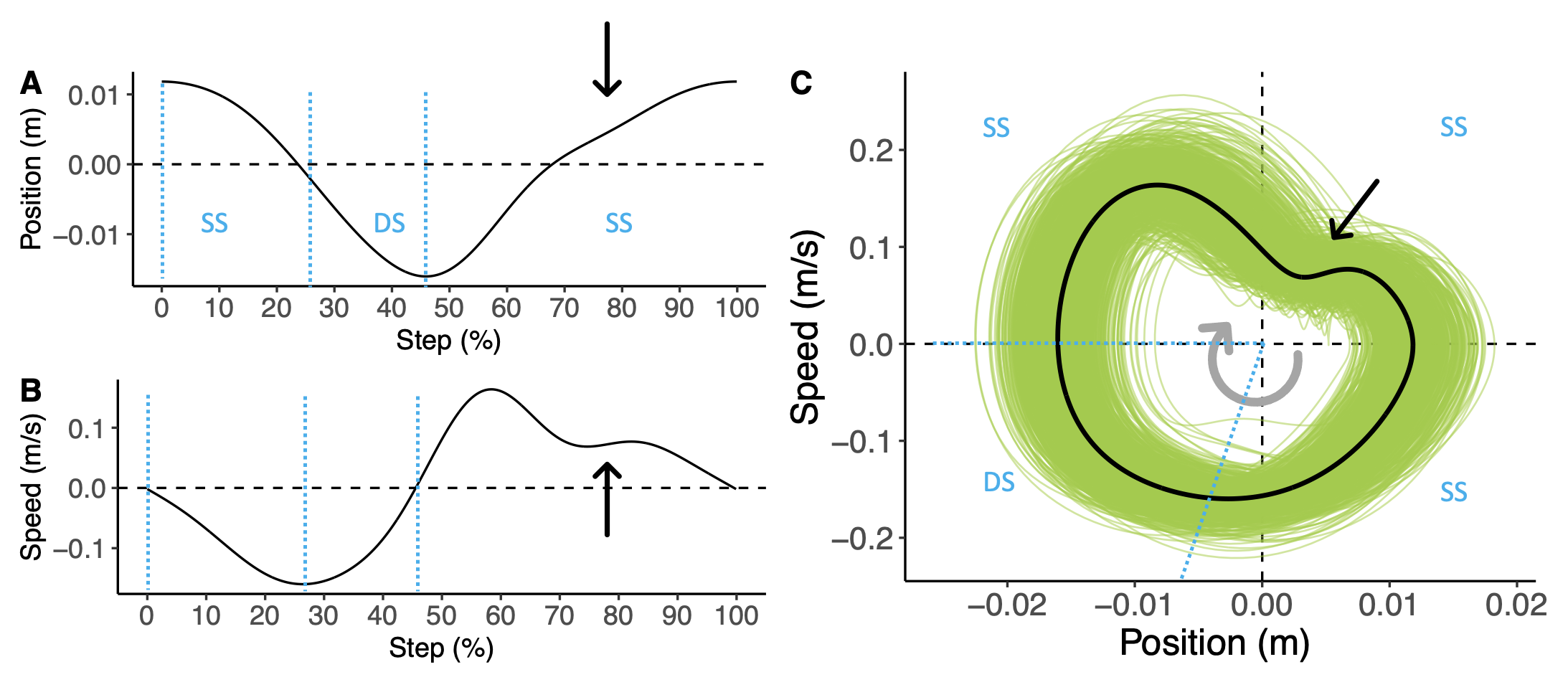}
	\caption{A: Attractor of the centred vertical position of the COM versus time, expressed in \% of step time. B: Attractor of the vertical speed of the COM versus time, expressed in \% of step time. C: Typical plot of COM vertical trajectory in phase space (solid green lines representing 508 gait cycles) during walking for a participant and of the corresponding attractor (solid black line). The straight arrows outline the  deviation from genuine harmonic oscillator. The curved arrow is the arrow of time. Note that a closed loop corresponds to one step cycle, a complete gait cycle being composed of two step cycles. The blue dotted line separate the single stance (SS) and dual stance (DS) phases.}
	\label{fig0}  
\end{figure}

The attractors of the centred vertical position and speed of the COM versus time are shown in 
Fig. \ref{fig0} A and B, and a typical phase space trajectory is also shown in Fig. \ref{fig0} C. The 
attractor is computed as follows. After each step cycle is identified, an average cycle is computed. For 
this purpose, each step was normalised to a duration of 1 time unit (0--100\%). Then 1200 bins, one for 
each frame, were created and filled with the data of all steps of a given participant under a given 
condition. For each bin, the mean and standard deviation were computed. %The mean time series are shown in Figs. \ref{fig0} A and B.
This yields the average cycle, which we refer to as the attractor, following works such as those of 
\cite{BROSCHEID2018242,Raffalt}. The attractor may be interpreted as the basic motor pattern that a 
participant tries to achieve during each step cycle -- without achieving it exactly due to intrinsic 
physiological noise. 

From the attractor, it is easy to see that the effective dynamics is not a pure harmonic oscillator, 
as it moves away from an elliptical shape in the first quadrant 
(indicated by a straight arrow in Fig. \ref{fig0} C).\footnote{We use the trigonometric convention 
in order to split the plane into four quadrants, with the angle going from $0$ to $\pi/2$ in the first quadrant, 
from $\pi/2$ to $\pi$ in the second quadrant, etc.}
The deformation is systematic and present in all participants. Therefore, the model presented in section 
\ref{diffusion} may be applied since the diffusion coefficient can be nonzero. Here, each step cycle starts 
when the COM is at its higher position and its speed is null, i.e., when the subject is in midstance: one foot 
on the ground, the knee is extended and the other foot is in swing phase and crossing the stance leg. The 
direction of the trajectory of the COM in phase space is clockwise: from fourth to first quadrant. 
In the fourth and third quadrants, the COM position decreases (downward movement) 
and the speed is negative. The attractor shape is elliptical as in a spring-mass model 
of the stance leg \cite{spring}, inducing a harmonic motion. 
In the second and first quadrants, the COM position increases (upward movement) and 
its speed is now positive. In the fourth quadrant, the participant is in single leg stance (SS) on one foot and 
this phase continues during the first part of the third quadrant. In the second part of the third quadrant, the 
participant is in dual stance (DS), that begins when the COM speed is at its lowest value and ends when 
the COM postion is at its lowest value \cite{Adamczyk}. At the end of the second quadrant and the first 
one, the participant is in single leg stance on the other foot.

%\begin{table}[H]
%	\caption{Results of the fits of experimental distributions of the adiabatic invariants to model (\ref{solu}). Results are written under the form  median [Q1-Q3]. The $p-$values of the one-sample t-tests are given in the last column.}
%	\centering
	%% \tablesize{} %% You can specify the fontsize here, e.g., \tablesize{\footnotesize}. If commented out \small will be used.
%	\begin{tabular}{ccc}
%		\toprule
%		$D$ (10$^{-9}$ m$^2$/s) & 	6.26 [4.72$-$15.6] & $0.003$ \\
%		$\pi\ I_0$ (J.s/kg) & 0.0129 [0.0105$-$0.0178]	& $<0.001$  \\
%		$\delta$ (\%) &	2.63 [$-$4.59, 7.81] & 0.947 \\
%		\bottomrule
%	\end{tabular}\label{tab2a}
%\end{table}

\begin{table}[h]
	\caption{Results of the fits of experimental distributions of the adiabatic invariants to model (\ref{solu}). Results are written under the form  median [Q1$-$Q3]. The $p-$values of the one-sample t-tests are given in the last column.}
	\centering
	%% \tablesize{} %% You can specify the fontsize here, e.g., \tablesize{\footnotesize}. If commented out \small will be used.
	\begin{tabular}{ccc}
\hline
		$D$ (10$^{-9}$ m$^2$/s) & 11.618 [6.024$-$37.712] & <0.001 \\
		$\pi\ I_0$ (J.s/kg) & 0.0123 [0.0061$-$0.0178]&  <0.001 \\
		$\Pi$ (\%) & 100 [98.6-100] & \\
\hline
	\end{tabular}\label{tab2}
\end{table}

\begin{figure}[h]
	
	\centering
	\includegraphics[width=8cm]{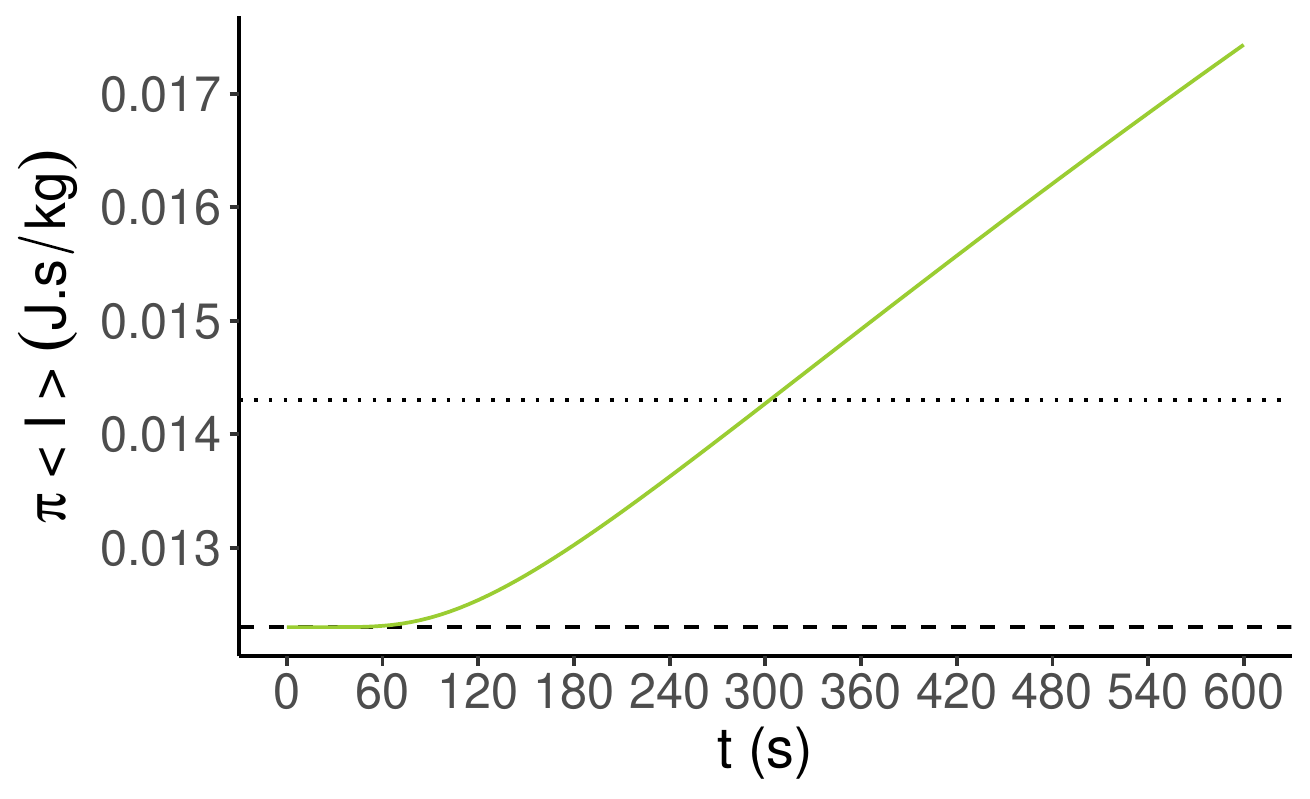}
	\caption{Plot of $\pi\, \left\langle I\right\rangle $ versus time (green line). The median value given in Table \ref{tab2} is indicated (dashed line), as well as the average value found in \cite{AdiabaticGait} (dotted line).}
	\label{figav}  
\end{figure}

It appears that the fit is relevant since $\Pi>\,$97\% for 20 participants out of 25. 
Hence, the model (\ref{solu}) 
fairly well agrees  with the time evolution of the distribution 
of the adiabatic invariant. Fitted parameters are summarized in Table \ref{tab2}. %The value  $\delta=0$ is compatible with our results, which is coherent with the zero-drift approximation made in this study. 
The mean value of $I$ reads 
	\begin{equation}
	\left\langle I\right\rangle=\int^{+\infty}_0 \rho(I,t)dI=\frac{\Theta(t)}{{\rm erf}\left(\frac{I_0}{\sqrt{4Dt}} \right)}\ I_0\;,
	\end{equation}
and its behaviour versus time is displayed in Fig. \ref{figav}. The mean value stays of order $I_0$ during the protocol: Less than 10 \% of variation is observed. The values obtained are comparable to the mean value found by an independent analysis in \cite{AdiabaticGait}: $\pi\, I=$0.0143$\pm$0.0058 J.s/kg.

\begin{figure}[h]

	\centering
	\includegraphics[width=15cm]{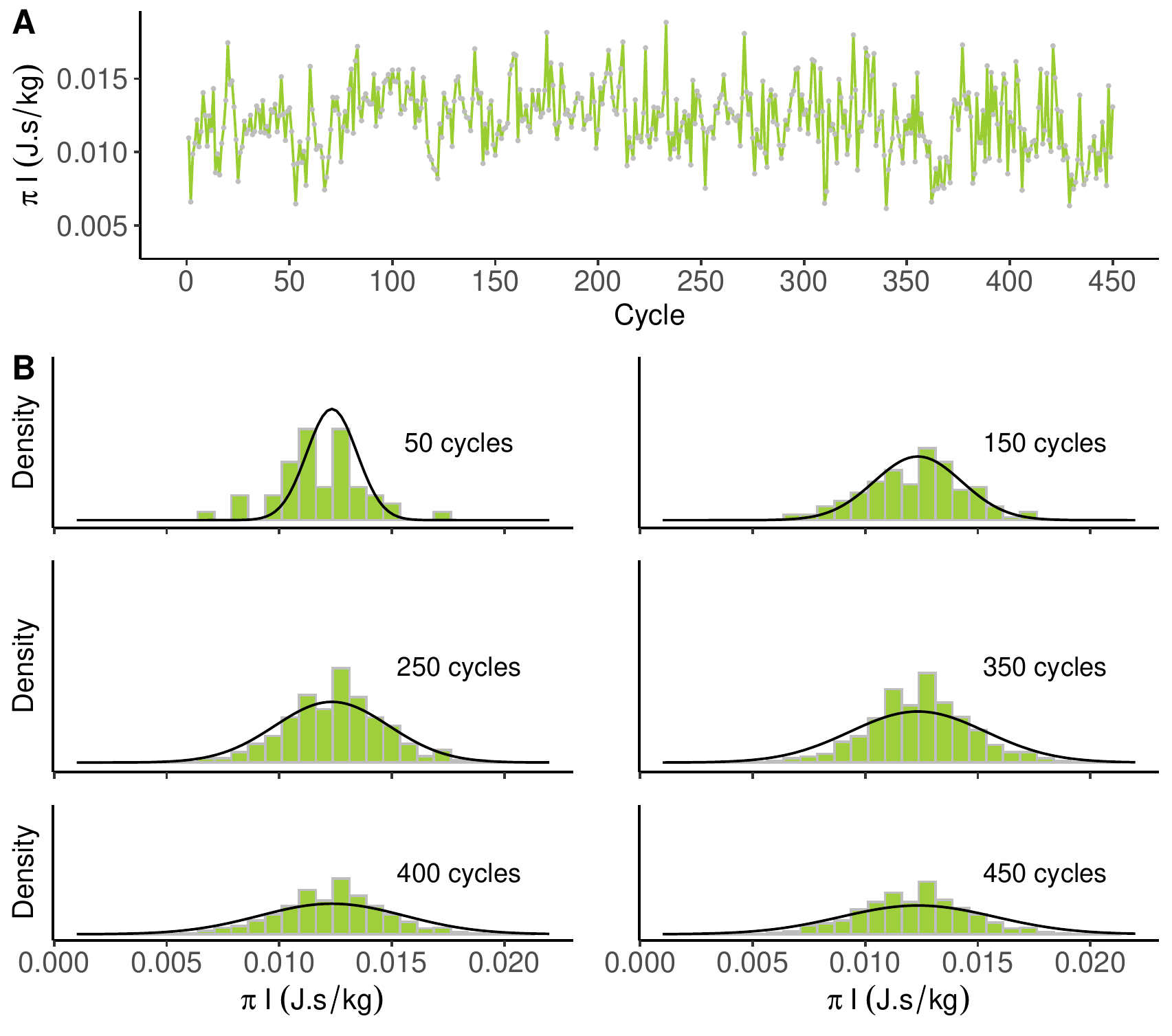}
	\caption{A: Adiabatic invariant versus time (gray points) for a given participant. Lines are added to guide the eyes, and time is expressed in cycle number. B: Typical plots showing the comparison between the theoretical distribution $\rho^{{\rm th}}(t_i,I)$ (solid line) and the experimental  one $\rho^{{\rm exp}}(t_i,I)$ (histograms) after 50, 150, 250, 350, 400 and 450 cycles for the same participant as in A, with $\Pi=99.4$\%. Fitted parameters are equal to $\pi\, I_0=0.0123$ J.s/kg and $D=1.05\, 10^{-8}$ m$^2$/s. }
		\label{fig}  
\end{figure}

 The ability of the model to fit the data can be appraised in Fig. \ref{fig}, where a typical plot of the fitted distributions versus experimental observations is displayed for one participant. All participants show the same qualitative agreement between the model and the data. 

%\begin{listing}[H]
%\caption{Title of the listing}
%\rule{\textwidth}{1pt}
%\raggedright Text of the listing. In font size footnotesize, small, or normalsize. Preferred format: left aligned and single spaced. Preferred border format: top border line and bottom border line.
%\rule{\textwidth}{1pt}
%\end{listing}

%%%%%%%%%%%%%%%%%%%%%%%%%%%%%%%%%%%%%%%%%%
\section{Discussion}\label{conclu}

By studying the vertical motion of the healthy participant during walking, we have shown that the phenomenon of phase space diffusion can be observed through the distribution of adiabatic invariant values over time. To our knowledge, this is the first time that such an observation is made in human motion. 

The time evolution of the distribution of the adiabatic invariant over time is compatible with the Fokker-Planck equation with constant diffusion coefficient for healthy young adults walking at spontaneous speed of progression. Thus, up to our experimental precision, we observe no drift and no deformation of the constant-$D$ distribution for high or low values of $I$. A change in the most likely value of $I$ can presumably be associated with a change in energy expenditure during walking. As argued in \cite{AdiabaticGait}, the value of the adiabatic invariant should be proportional to oxygen consumption during walking, and an increase in the former should be associated with an increase in the latter.  

There are a number of immutable factors in the environment in which we live. One is gravity. The brain, instead of fighting against the effects caused by its presence (e.g. the emergence of a weight that counteracts movements) has developed strategies to make the most of it and optimize movements \cite{white20}. In other words, humans move more optimally in the presence than in the absence of a gravitational field. We can make a parallel with the existence of noise in physiological systems. These emerge at every level of the decision-action chain, from perception to motoneurons. Authors have proposed in the optimal movement variability framework, that the central nervous system could actually exploit the presence of noise and hence, act more optimally in the presence of certain levels of uncertainties. Following \cite{Goldberger,stergiou2006,VANEMMERIK20163} we interpret the variability measured by the distribution not as an ``imperfection'' but rather as an indication of the adaptability of the participants to the motor task. A given value of the adiabatic invariant corresponds to a given area in phase space for the step cycle under consideration. Thus, the changes in $I$ indicate that the participants have access to a wide range of motor strategies, visualised as closed step cycles in phase space. The distribution becomes wider and wider over time: more and more different motor patterns are ``explored''. In our approach, there is no drift: the most likely value of $I$, i.e. the attractor defined as the ideal trajectory in phase space that the participant is aiming for, does not change with time. 

\begin{figure}[h]
	
	\centering
	\includegraphics[width=8cm]{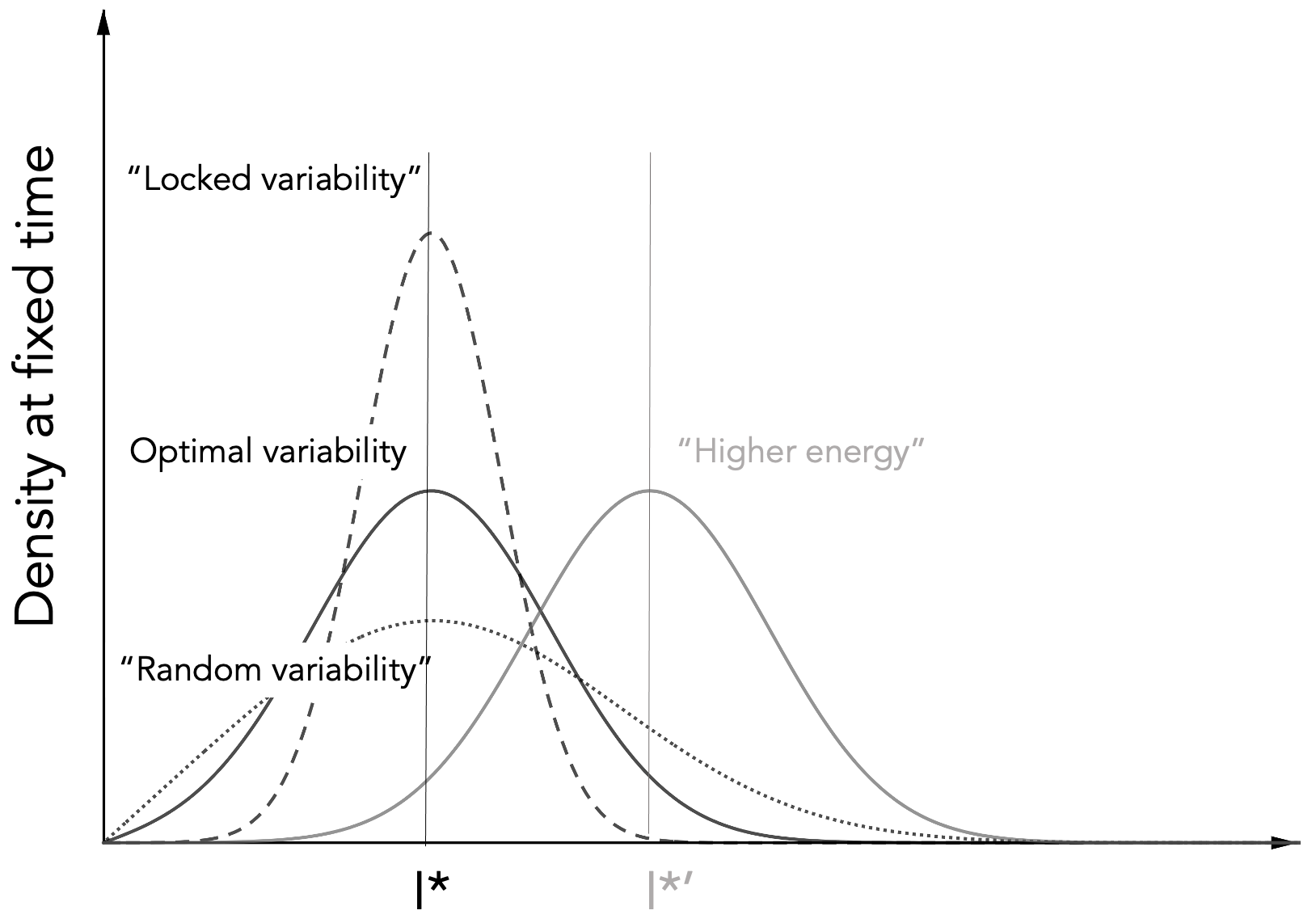}
	\caption{Schematic representation of several types of adiabatic invariant densities. We assume that the black solid line is the density of a young, healthy, individual. The ``locked'' (dashed line) and ``random'' (dotted line)  curves correspond to, respectively, smaller and higher diffusion coefficients than the optimal one. The ``higher energy'' curve (gray solid line) has an optimal diffusion coefficient but a higher maximally probable $I^*$, denoted $I^*$' on the horizontal axis.}
	\label{fig2}  
\end{figure}

We conjecture that the shape of the distribution $\rho(I,t)$ might be sensitive to the experimental condition and/or to each participant, as shown in Fig. \ref{fig2}. In particular, there should be an optimal value for the diffusion coefficient $D$ and for $I_0$ for a young, healthy individual. Too large a value for $D$ would reflect a lack of or altered motor control of the participant, leading to variability that tends to be random, as observed in stride interval variability of patients with neurodegenerative diseases \cite{MOON2016197} for example. Too small a value for $D$ could be related to insufficient adaptability of the participant: the number of available patterns (i.e. different values of $I$) is not maximal. Such a case is observed, for example, in the electrocardiographic signal of patients with cardiovascular disease \cite{doi:10.1073/pnas.012579499} or in healthy children, whose walking patterns are more stereotyped than in adults \cite{hausdorff1999maturation}. The diffusion coefficient then offers a novel way to quantify the general behaviour of internal models developed for a given task. Indeed, wide distribution (high D) are observed after time spent to experience or explore a task. On the other hand, narrow distributions (small D) may reflect a lack of generalisation of the motor strategies adopted. In motor control – and rehabilitation in particular – the concept of generalisation is tightly linked to the one of transfer \cite{dizio,criscimagna,sarwary}. When working toward recovering lost or impaired motor functions, the challenge is to find the best possible movements that may be transferred to as many functional tasks as possible. These movements may be interpreted as fundamental bricks of the action repertoire. An interesting question is why would a participant opt for a narrow distribution? One possible explanation for this is related to the way motor learning works. There are different learning mechanisms, the most powerful being error-based learning. In this one, one plans the best possible action by minimising a cost function that includes target reaching in the general sense and effort. An error signal is observed in case of discrepancy between observation and what has been predicted by forward models \cite{shad94,Thoroughman2000} which induces strategic changes, and encourage exploration. Another learning mechanisms, however, co-exists, with a slower dynamics: use-dependent-learning. When relying on this mechanism, one tends to repeat the same action if it led to success in the past, thereby discouraging exploration in task space. Adopting this strategy results from a compromise between cost and benefit: the target may be reached, but the control policy may be stuck in a local minima of the cost function. 

We hope to apply the present formalism to participants with different ages or experimental conditions to investigate the effects of deviations from the optimal ``healthy young adult state'' on $\rho(I,t)$ in future work. More precisely, we hope to design appropriate experimental contexts that would manipulate $I$ and $D$ independently, then providing a better functional understanding of these indexes in motor control.

\medskip

\textit{Acknowledgements}

The authors thank G. Henry and F. Piccinin for data acquisition.

%%%%%%%%%%%%%%%%%%%%%%%%%%%%%%%%%%%%%%%%%%
%\reftitle{References}

\end{document}